\documentclass[
reprint,
amsmath,amssymb,
aps,
prb,
]{revtex4-2}

\usepackage{hyperref}
\hypersetup{colorlinks=true,
            linkcolor=black,
            citecolor=black,
            urlcolor=blue}
\usepackage{amssymb}
\usepackage{mathbbol}
\usepackage{courier}

\usepackage{graphicx}
\usepackage{dcolumn}
\usepackage{bm}

\begin{document}

\title{Yu-Shiba-Rusinov states in color superconducting quark matter}



\author{Virgil V. Baran$^{1,2,3}$}
\email[]{virgil.v.baran@unibuc.ro}
\author{Jens Paaske$^{3}$}
\affiliation{$^{1}$Faculty of Physics, University of Bucharest, 405 Atomi\c stilor, RO-077125, Bucharest-M\u agurele, Romania}
\affiliation{$^{2}$"Horia Hulubei" National Institute of Physics and Nuclear Engineering, 30 Reactorului, RO-077125, Bucharest-M\u agurele, Romania}
\affiliation{$^{3}$Center for Quantum Devices, Niels Bohr Institute, University of Copenhagen, 2100 Copenhagen, Denmark}

\begin{abstract}

The high-density region of the QCD phase diagram displays an intricate competition between color superconductivity and the QCD Kondo effect due to color exchange in quark matter containing a single heavy quark impurity. We explore the characteristic impurity-induced superconducting subgap states arising in such systems by generalizing the surrogate model solver, recently considered in the context of condensed matter physics [Phys. Rev. B 108, L220506 (2023)]. The method consists of approximating the full superconducting bulk by only a small number of effective levels whose parameters are chosen so as to best reproduce the Matsubara frequency dependence of the impurity-bulk hybridization function. We numerically solve a surrogate QCD Kondo model describing a quantum impurity color-exchange coupled to a two-color two-flavor superconducting bulk. The results directly indicate the presence of multiple phase transitions as the coupling of the impurity to the bulk is increased, due to the interplay between various overscreened states. The methods introduced here are straightforward enough to be extended to more realistic QCD scenarios.

\end{abstract}

\maketitle

\section{Introduction}

Heavy quarks are essential probes for the study of hot and dense matter arising in heavy-ion collisions and compact neutron or quark stars~\cite{Neubert1994Sep,Neubert1996Oct,Manohar2000Mar,Hosaka2017Sep}. In this context, the heavy charm or bottom quarks may be regarded as impurities immersed in a medium
composed of light up, down and strange quarks, with the complex non-Abelian interactions specific to quantum chromodynamics (QCD) then leading to a subtle interplay of various many-body effects. In particular, heavy quarks or hadrons may experience the Kondo effect through interactions (in the color or isospin channels) with
the nuclear or quark medium~\cite{Yasui2013Jul,Hattori2015Sep,Yasui2017Jul,Yasui2017Oct,Yasui2017Dec,Kimura2019Jan,Hattori2019May,Yasui2019Oct,Suenaga2021Mar,Ishikawa2021Nov,Suenaga2022Apr,Yasui2024May}. At low baryon densities, this QCD Kondo effect competes with the formation of the chiral condensate, and at high baryon densities with the formation of a diquark condensate~\cite{Suzuki2017Dec,Kanazawa2016Dec}. In the latter case, the Fermi surface is
subject to the Cooper instability due to the attractive
interactions between light quarks, which favor a color superconducting ground state at low temperatures~\cite{Rajagopal2001Apr, Alford2008Nov}.

The interplay of Kondo screening and superconductivity has been investigated for decades in condensed matter physics~\cite{Yu1965,Shiba1968,Rusinov1969,Takano1969Jan, Sakurai1970Dec,Muller-Hartmann1971Feb,Matsuura1977Jun,Cuevas2001Feb,Franke2011May, Zazunov2018Nov}, with the complex properties of the impurity models describing Kondo systems having been mapped out using a complementary set of techniques, such as the scaling
approach~\cite{Anderson1970Dec}, Wilson’s numerical renormalization group (NRG)~\cite{Wilson1975Oct}, the Bethe
ansatz solution~\cite{Andrei1980Aug,Vigman1980Apr, Andrei1983Apr, Tsvelick1983Jan}, and various large-N expansion schemes~\cite{Read1983Jun,Coleman1984Mar}. The interest in such problems has been revived in recent years following the discovery of the Kondo effect in quantum dots (QDs)~\cite{Goldhaber-Gordon1998Jan,Cronenwett1998Jul,Nygard2000Nov,Quay2007Dec,Jespersen2006Dec,Csonka2008Nov}, which essentially behave as tunable magnetic impurities. When tunnel-coupled to superconducting leads, they induce so-called Andreev bound states inside the superconducting gap. These subgap states range from Yu-Shiba-Rusinov (YSR) states~\cite{Yu1965, Shiba1968, Rusinov1969}, induced by odd occupied Coulomb blockaded QDs with a local magnetic moment, to a localized quasiparticle excitation above an induced gap on proximitized QDs with smaller charging energy~\cite{Bauer2007Nov, Meng2009Jun, Oguri2013, Kirsanskas2015}.

As the finite BCS gap of the superconducting quasiparticle excitations essentially cuts off all logarithmic singularities and hinders an actual Kondo problem, the calculational effort required to obtain the low-lying subgap bound states turns out to be rather modest. This is due to the smooth and well-behaved dependence of the BCS Green's functions on the Matsubara frequency, which admits a rapidly converging few-level approximation. The  latter allowed us to match the numerically exact single-impurity NRG results with a low-dimensional surrogate BCS model and to then explore the subgap physics of more complicated multi-impurity and/or multi-terminal problems \cite{Baran_PRB2023, Baran2024Jun, Baran2024Aug}.  

Impurity problems of similar complexity naturally arise when considering the extra color and flavor degrees of freedom inherent to QCD. The main purpose of this work is thus to show that the above surrogate model solver (SMS) is an efficient method for studying the low-lying states induced by quantum impurities in high-density color superconducting matter.  This constitutes an unbiased approach that goes beyond the limiting classical spin approximation \cite{Kanazawa2016Dec}, and is thus able to account for the various phase transitions expected for an impurity with multiple quantum numbers, as shown below. Furthermore, it provides a complementary perspective to mean-field based studies of the QCD Kondo effect, see e.g. the recent Ref.~\onlinecite{Yasui2024May} and references therein, and to recent exact solutions \cite{Kattel2024Dec}.

The paper is organized as follows. In Sec.~\ref{sec:modeling} we define the  single-impurity QCD Kondo model and to which we subsequently apply the SMS approximation. In Sec.~\ref{sec:results}, we outline and comment on the numerical results regarding the phase diagrams of the surrogate Kondo models and their corresponding correlation functions. The final Sec.~\ref{sec:conclusions} is devoted to the conclusions and perspectives.

\section{Modeling methodology}
\label{sec:modeling}

\subsection{Color superconducting Kondo model}

Following Ref.~\onlinecite{Kanazawa2016Dec}, we consider a toy model of two-color two-flavor Dirac fermions
\begin{equation}
\begin{aligned}
\label{h0}
\hat{\mathcal{H}}_0=\int\text{d}^3x\{\psi^\dagger(\text{i}v{\boldsymbol{\sigma}}\cdot\nabla-\mu)\psi-\frac{\Delta}{2}[\psi^T\sigma_2 \tau_2 t_2\psi +\text{h.c.}]\},
\end{aligned}
\end{equation}
suitable to describe the two-flavor color-superconducting phase that may be realized in the interior of compact stars \cite{Alford1998Mar,Rapp1998Jul}. In this phase, quarks with one out of three colors remain gapless and neutral under the unbroken gauge group SU(2) $\subset$ SU(3), their interactions
with the gapped quarks and impurities being suppressed at low energies. Thus we are led to the above model involving two-color two-flavor Dirac fermions with a Majorana mass $\Delta$. As in Ref.~\onlinecite{Kanazawa2016Dec}, we consider for simplicity a single quark chirality.

However, while Ref.~\onlinecite{Kanazawa2016Dec} relies on the classical spin method for extracting the subgap spectrum, we consider here a quantum impurity color-exchange coupled to the bulk medium,
\begin{equation}
\label{hexc}
    \hat{\mathcal{H}}_\text{exc}=J_0\, \boldsymbol{T}_{\text{imp}}\cdot \psi^\dagger(\textbf{0}) \,\boldsymbol{t}\,\psi(\textbf{0})~,
\end{equation}
leading to the total QCD Kondo model Hamiltonian
\begin{equation}
\label{htot}
    \hat{\mathcal{H}}_{\text{Kondo}}=\hat{\mathcal{H}}_0+\hat{\mathcal{H}}_\text{exc}~.
\end{equation}

Above, the light-quark field $\psi$ carries
spin $(\uparrow,\downarrow)$, color $(\Uparrow,\Downarrow)$ and flavor $(u,d)$ quantum numbers and $\boldsymbol{T}_{\text{imp}}$ represents the impurity's color pseudo-spin with magnitude $T=1/2$. Note that while the bulk Hamiltonian $\hat{\mathcal{H}}_0$ of Eq.~(\ref{h0}) is invariant under the SU(2)$\times$SU(2)$\times$SU(2) that acts on the spin($\sigma)$-flavor($\tau$)-color($t$) indices, the spin of the impurity quark is absent from $\hat{\mathcal{H}}_{\text{Kondo}}$ as heavy quarks in quark matter do not experience spin-dependent interactions \cite{Neubert1994Sep,Manohar2000Mar,Isgur1990Mar}.

Eq.~(\ref{h0}) may be rewritten in the Nambu basis $\Psi_{\bf{k}}=(\psi_{\bf{k}}~\psi_{-\bf{k}})^T$ as 
\begin{equation}
\label{h0nambu}
    \hat{\mathcal{H}}_0=\frac{1}{2}\int\frac{\text{d}^3\bf{k}}{(2\pi)^3}~\Psi_{\bf{k}}^\dagger \mathcal{H}_{\bf{k}}\Psi_{\bf{k}}~,
\end{equation}
with
\begin{equation}
    \mathcal{H}_{\bf{k}}=\begin{pmatrix}
        (-v {\bf{k}}\cdot\boldsymbol{\sigma}-\mu)\otimes \mathbb{1}_4 & -\Delta \sigma_2 \tau_2 t_2\\
        -\Delta \sigma_2 \tau_2 t_2 & (-v {\bf{k}}\cdot\boldsymbol{\sigma}^T+\mu)\otimes \mathbb{1}_4
    \end{pmatrix}~,
\end{equation}
where $\mathbb{1}_4=\mathbb{1}_2^{\text{color}}\otimes \mathbb{1}_2^{\text{flavor}}$. In the following, we will need the momentum integrated Green's function for $G_0({\bf{k}},\text{i}\omega)=1/(\text{i}\omega-\mathcal{H}_{\bf{k}})$ which will be reinterpreted below Eq.~(\ref{tunneling_self_en}) as the tunneling self-energy $\Sigma_T$,
\begin{equation}
\label{tunnelselfenergy}
\begin{aligned}
\Sigma_T\equiv J_0\int\frac{\text{d}^3\bf{k}}{(2\pi)^3}G_0({\bf{k}},\text{i}\omega)=J_0\int\frac{\text{d}^3\bf{k}}{(2\pi)^3}\frac{1}{\mathcal{X}^2-4\mu^2v^2k^2}\\
\begin{pmatrix}
        [(\text{i}\omega-\mu)\mathcal{X}-2\mu v^2k^2] \mathbb{1}_8 & -\Delta \sigma_2 \tau_2 t_2 \mathcal{X}\\
        -\Delta \sigma_2 \tau_2 t_2 \mathcal{X}& [(\text{i}\omega+\mu)\mathcal{X}+2\mu v^2k^2] \mathbb{1}_8 
    \end{pmatrix}~,
\end{aligned}
\end{equation}
where $\mathcal{X}=-\omega^2-\mu^2-\Delta^2-v^2k^2$. For a weak superconductor $\Delta\ll \mu$, this simplifies to $\mathcal{X}\approx-\mu^2-v^2k^2$. Furthermore, if we restrict the domain of integration to a thin shell of half-bandwidth $D$ around the Fermi surface as in the conventional BCS theory, we get $\mathcal{X}\approx-2\mu^2$ and
\begin{equation}
\begin{aligned}
\Sigma_T=\mathcal{J} 
\begin{pmatrix}
        -\text{i}\omega \mathbb{1}_8 & \Delta \sigma_2 \tau_2 t_2\\
        \Delta \sigma_2 \tau_2 t_2 & -\text{i}\omega \mathbb{1}_8
    \end{pmatrix}\,g(\omega)~,
\end{aligned}
\end{equation}
with 
\begin{equation}
\label{gfunction}
    g(\omega)\equiv \frac{1}{\pi}\int_{-D}^D\text{d}\xi\,\frac{1}{\xi^2+\Delta^2+\omega^2}=\frac{2}{\pi}\frac{ \arctan\left(\frac{D}{\sqrt{\Delta^2+\omega^2}}\right)}{\sqrt{\Delta^2+\omega^2}}~,
\end{equation}
where we denoted the dimensionless color-exchange coupling by $\mathcal{J}=\pi\rho(\mu)J_0/2$, which is given in terms of the  density of states at the Fermi energy in the normal state, $\rho(\mu)=\mu^2/(2\pi^2v^3)$. Let us note that the thin shell approximation only introduces mild quantitative effects for the resulting phase diagram \cite{Kanazawa2016Dec}. We thus leave the rigorous treatment of the above Eq.~(\ref{tunnelselfenergy}) for future works devoted to more realistic applications.

\subsection{Surrogate Kondo models}

We now proceed to formulate the SMS prescription for generic superconducting Kondo models. The presence of fermionic bilinears in the exchange term of Eq.~(\ref{hexc}) prevents us from directly applying the SMS methodology as in the case of Anderson impurities linearly (tunnel) coupled to the superconducting bulk \cite{Baran_PRB2023}. We may nevertheless consider the partition function corresponding to Eq.~(\ref{htot}) for the bulk fermions with a frozen impurity configuration, expressed as the coherent state path integral
\begin{align}
Z[\boldsymbol{T}_{\text{imp}}]=\!\int\!\mathcal{D}\psi~e^{J_0\, \boldsymbol{T}_{\text{imp}}\cdot (\sum_{\bf{k}}\psi^\dagger_{\bf{k}})\boldsymbol{t}(\sum_{{\bf{k}'}}\psi_{\bf{k}'})+\sum_{\bf{k}}\psi^\dagger_{\bf{k}} G_0^{-1}\psi_{\bf{k}}},
\end{align}
where the summation over Matsubara frequencies is implied. The exchange coupling term may be eliminated by a Hubbard-Stratonovich transformation introducing a local auxiliary $\eta$ field, and subsequently integrating out the $\psi$ fermions 
allows us to express the partition function as 
\begin{align}
\label{tunneling_self_en}
Z[\boldsymbol{T}_{\text{imp}}]=\!\int\!\mathcal{D}\eta~e^{\eta^\dagger (\boldsymbol{T}_{\text{imp}}\cdot \boldsymbol{t})^{-1} \eta+\eta^\dagger \Sigma_T \eta}~,
\end{align}
written in terms of the tunneling self-energy defined in Eq.~(\ref{tunnelselfenergy}) above.

At this point, the $g$-function of Eq.~(\ref{gfunction}), and thus the tunneling self-energy itself, of the full continuous BCS model may be approximated within the SMS approach \cite{Baran_PRB2023} by that corresponding to a small number $\tilde{L}$ of effective levels
\begin{equation}
\label{tildeg}
\begin{aligned}
    &\tilde{g}_{\text{even}}(\omega)\equiv2\sum_{\ell=1}^K  \frac{\gamma_\ell}{\tilde\xi_\ell^2+\Delta^2+\omega^2}~, ~\tilde L =2K~,\\
    &\tilde{g}_{\text{odd}}(\omega)\equiv\frac{\gamma_0}{\Delta^2+\omega^2}+\tilde{g}_{\text{even}}(\omega)~,~\tilde L =2K+1~ .\\
    \end{aligned}
\end{equation}

Note that each SC level with energy $\xi$ contributes with a factor of $(\xi^2+\Delta^2+\omega^2)^{-1}$ to the $g$-function of Eq.~(\ref{gfunction}).
A $\tilde g$-function as in Eq.~(\ref{tildeg}) may thus be obtained by integrating out an effective superconductor with the same gap, $\Delta$, as the original one and whose $\tilde L$ discrete levels with energies $\pm|\tilde\xi_\ell|$ are coupled to the $\eta$ field via tunneling matrix elements $\tilde t_\ell\sim\sqrt{\gamma_\ell}$. The parameters $\{\gamma_\ell,\tilde{\xi}_\ell\}$ define the surrogate model Hamiltonian as a discretized version of Eq.~(\ref{h0nambu}). They are determined in Ref.~\onlinecite{Baran_PRB2023} by minimizing the cost function $ \chi^2=\sum_j |g(\omega_j) - \tilde g(\omega_j)|^2$, evaluated on a logarithmically spaced grid of frequencies $\omega_j$ in order to ensure a good fit at subgap frequencies. The numerical values used in this work are the same as in Ref.~\onlinecite{Baran_PRB2023}  and  correspond to $D=10\Delta$, namely $\gamma_0=0.9618$ for $\tilde{L}=1$, $\gamma_1=1.2642$, $\tilde{\xi}_1=1.3099$ for $\tilde{L}=2$ and $\gamma_0=0.508$, $\gamma_1=1.8454$ and $\tilde{\xi}_1=2.7546$ for $\tilde{L}=3$ (all in units of $\Delta$).

Having obtained a suitable surrogate representation of the tunneling self-energy, we may now retrace our steps through the path integrations and arrive at the surrogate Kondo model
\begin{equation}
\label{htilde}
    \widetilde{\mathcal{H}}_{\text{Kondo}}=\widetilde{\mathcal{H}}_0+\widetilde{\mathcal{H}}_\text{exc}~,
\end{equation}
where
\begin{equation}
   \widetilde{\mathcal{H}}_0=\sum_{\ell=1}^{\tilde{L}}\tilde\xi_\ell \tilde{c}_\ell^\dagger \tilde{c}_\ell-\frac{\Delta}{2}(\tilde{c}_\ell^T\sigma_2 \tau_2 t_2\tilde{c}_\ell +\text{H.c.})~, 
\end{equation}
and
\begin{equation}
    \widetilde{\mathcal{H}}_\text{exc}=\mathcal{J} \,\boldsymbol{T}_{\text{imp}}\cdot\sum_{\ell,\ell'=1}^{\tilde{L}} \sqrt{\gamma_\ell \gamma_{\ell'}}\,\tilde{c}_\ell^\dagger\,\boldsymbol{t}\,  \tilde{c}_{\ell'}~.
\end{equation}
Here we suppressed the spin-color-flavor indices for the $\tilde{c}_\ell$ operators. Explicitly, $\tilde{c}_{\ell; \sigma\tau f}$ would create a fermion of spin $\sigma=\uparrow,\downarrow$, color $\tau=\Uparrow,\Downarrow$ and flavor $f=u,d$ on the level $\ell$. Explicitly, for the simplest $\tilde{L}=1$ or zero-bandwidth (ZBW) case the pairing term reads
\begin{equation}
   \widetilde{\mathcal{H}}_0^{(\tilde{L}=1)}=-\text{i}\Delta u_\Uparrow^\uparrow d_{\Downarrow}^{\downarrow}+\text{i}\Delta u_{\Uparrow}^{\downarrow}d_{\Downarrow}^{\uparrow}+\text{i}\Delta u_{\Downarrow}^{\uparrow}d_{\Uparrow}^{\downarrow}-\text{i}\Delta u_{\Downarrow}^{\downarrow}d_{\Uparrow}^{\uparrow}~ +\text{H.c.}
\end{equation}

Before discussing the results obtained by solving the above surrogate Kondo models, let us note that by increasing the number of surrogate levels $\tilde{L}$, the errors in the $g$-function fit are rapidly and systematically reduced by several orders
of magnitude across (and beyond) the fitting range. This ensures a fast convergence of the SMS subgap spectra and related observables (assuming the ratio of the SC gap to the Kondo temperature is reasonably small). Finally, we refer the reader to the Appendix for a short discussion on the simpler surrogate Kondo models for standard BCS problems.

\section{Numerical results}
\label{sec:results}

We solved for the lowest lying states in each fermion-parity sector of the surrogate Kondo models with $\tilde{L}=1$ by exact diagonalization, and for $\tilde{L}=1,2,3$ by the density matrix renormalization group (DMRG) in the matrix-product-state formulation, which is straightforward to implement with the ITensor library~\cite{White1992Nov, Schollwock2011Jan,itensor,itensor-r0.3}. For the DMRG calculations, we employed a maximum bond dimension of 5000 and an energy convergence threshold of $10^{-4}\Delta$.

\begin{figure}[t!]
\includegraphics[width=\columnwidth]{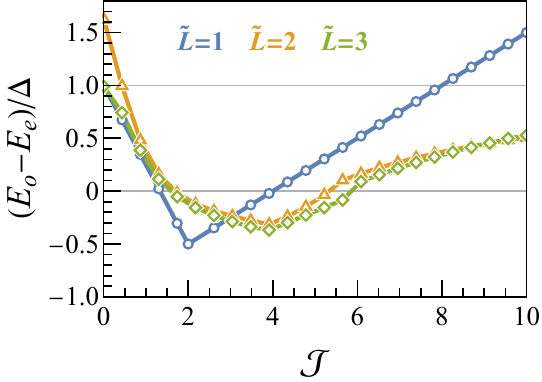}
    \caption{Energy difference $E_{o}-E_e$ (in units of $\Delta$) between the lowest even- and odd-fermion-parity states of the surrogate Kondo models for $\tilde{L}=1,2,3$ vs the dimensionless color-exchange coupling $\mathcal{J}$.}
\label{fig_1}
\end{figure}

The results regarding the evolution of the lowest lying subgap states in each fermion-parity sectors are shown in Fig.~\ref{fig_1}. First, all surrogate models agree on the crucial qualitative features of the phase diagram, directly indicating the existence of two distinct phase transitions with increasing coupling strength. From a more quantitative perspective, we notice some minor shortcomings in both the $\tilde{L}=1$ and $\tilde{L}=2$ surrogates. On the one hand, the overestimation of the strong-coupling excitation energy in the $\tilde{L}=1$ model is encountered also in the standard BCS case \cite{Baran_PRB2023}, and is related by the inability of the simple ZBW approximation (and, to a lesser extent, of the other odd-$\tilde{L}$ models) to properly screen the impurity. On the other hand, the $\tilde{L}=2$ model overestimates the excitation energy at weak couplings; this is caused by the absence of the $\tilde{\xi}_0=0$ level which would normally accomodate the lowest-lying screening quasiparticle at $E_{\text{qp}}=\Delta$, as is the case in the odd-$\tilde{L}$ models. Nevertheless, the $\tilde{L}=2,3$ models agree to a very good extent over the entire range of moderate to strong couplings shown in Fig.~\ref{fig_1}, thus validating the SMS convergence for the subgap spectrum. Even if naively one could argue that  smaller values of $\tilde{L}$ than in the standard BCS case are needed to achieve a good convergence (see. Ref.~\cite{Baran_PRB2023} and also the Appendix), the complexity of the surrogate model space also grows faster in the present case due to the extra color-flavor quantum numbers. 

\begin{figure}[t!]
\includegraphics[width=\columnwidth]{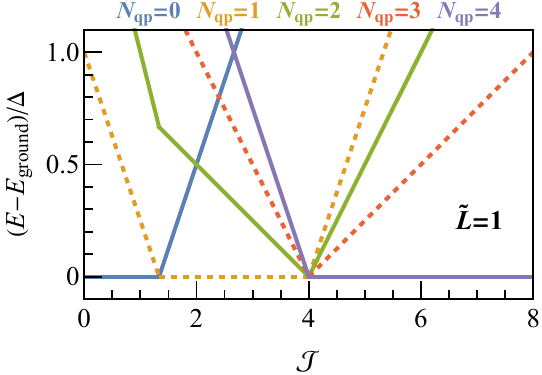}
    \caption{Energies (relative to the ground state) of the lowest-lying states in each sector of definite SC quasiparticle number $N_{\text{qp}}=0,...,4$ vs the dimensionless color-exchange coupling $\mathcal{J}$, as obtained by the $\tilde{L}=1$ (ZBW) model. Continuous (dashed) lines denote states with even (odd) fermion parity).}
\label{fig_2}
\end{figure}

To gain more insight into the impurity screening processes across the phase diagram, we plot in Fig.~\ref{fig_2} the evolution of the lowest-lying states in each sector of definite SC quasiparticle number $N_{\text{qp}}=0,...,4$ for the ZBW ($\tilde{L}=1$) model. In this simple case, their energies $E_{N_{\text{qp}}}$ may be computed analytically: 
$E(N_{\text{qp}})/\Delta=N_{\text{qp}}(1-\mathcal{J}/4)-\mathcal{J}/2$ for $N_{\text{qp}}\geq 1$, all relative to the unscreened state's constant reference value $E(N_{\text{qp}}=0)=0$.

As expected, the first phase transition is due to a single quasiparticle excitation becoming favored, with increasing coupling, over the SC ground state with no quasiparticles. Note that the former is a fourfold degenerate state due to the screening of the impurity's color pseudo-spin, say $\Uparrow$, by each of the four spin-flavor combinations $u_{\Downarrow}^{\uparrow},u_{\Downarrow}^{\downarrow},d_{\Downarrow}^{\uparrow},d_{\Downarrow}^{\downarrow}$. At larger couplings, the states with extra screening quasiparticles become increasingly favored. Incidentally, the present symmetric coupling situation features the special point $\mathcal{J}=4$ where all quasiparticle excitations become degenerate. Beyond this value, the ground state will involve all four spin-flavor combinations screening the impurity's color.

Overall, the ZBW ground state displays an $N_\text{qp}=0\rightarrow 1\rightarrow 4$ evolution with increasing coupling, lacking the phases with two or three screening quarks which only appear as parity flipping  excitations in the subgap spectrum.

\begin{figure}[t!]
\includegraphics[width=\columnwidth]{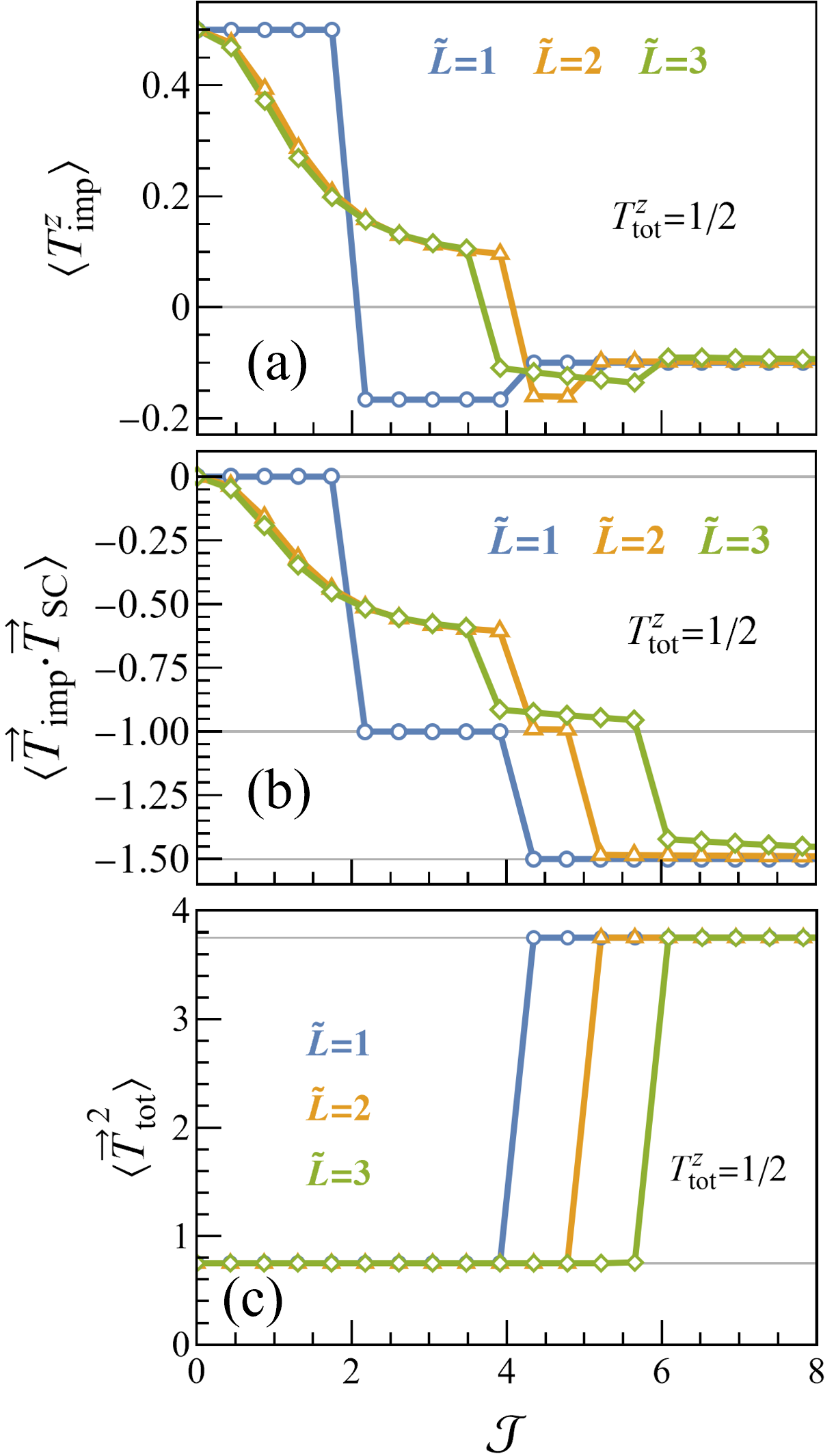}
    \caption{Color pseudo-spin properties of the lowest $T^z_{\text{tot}}=1/2$, even fermion-parity states vs the dimensionless color-exchange coupling $\mathcal{J}$, for the surrogate Kondo models $\tilde{L}=1,2,3$. (a)   
    Average impurity color pseudo-spin $\langle T_{\text{imp}}^z\rangle$. (b) Impurity-bulk color pseudo-spin correlation function $\langle \vec{T}_{\text{imp}}\cdot \vec{T}_{\text{SC}}\rangle$, with gridlines at $0, -1, -1.5$. (c) Total color pseudo-spin squared $\langle \vec{T}_{\text{tot}}^2\rangle$, with gridlines at $0.75$ and $3.75$.}
\label{fig_3}
\end{figure}

\begin{figure}[t!]
\includegraphics[width=\columnwidth]{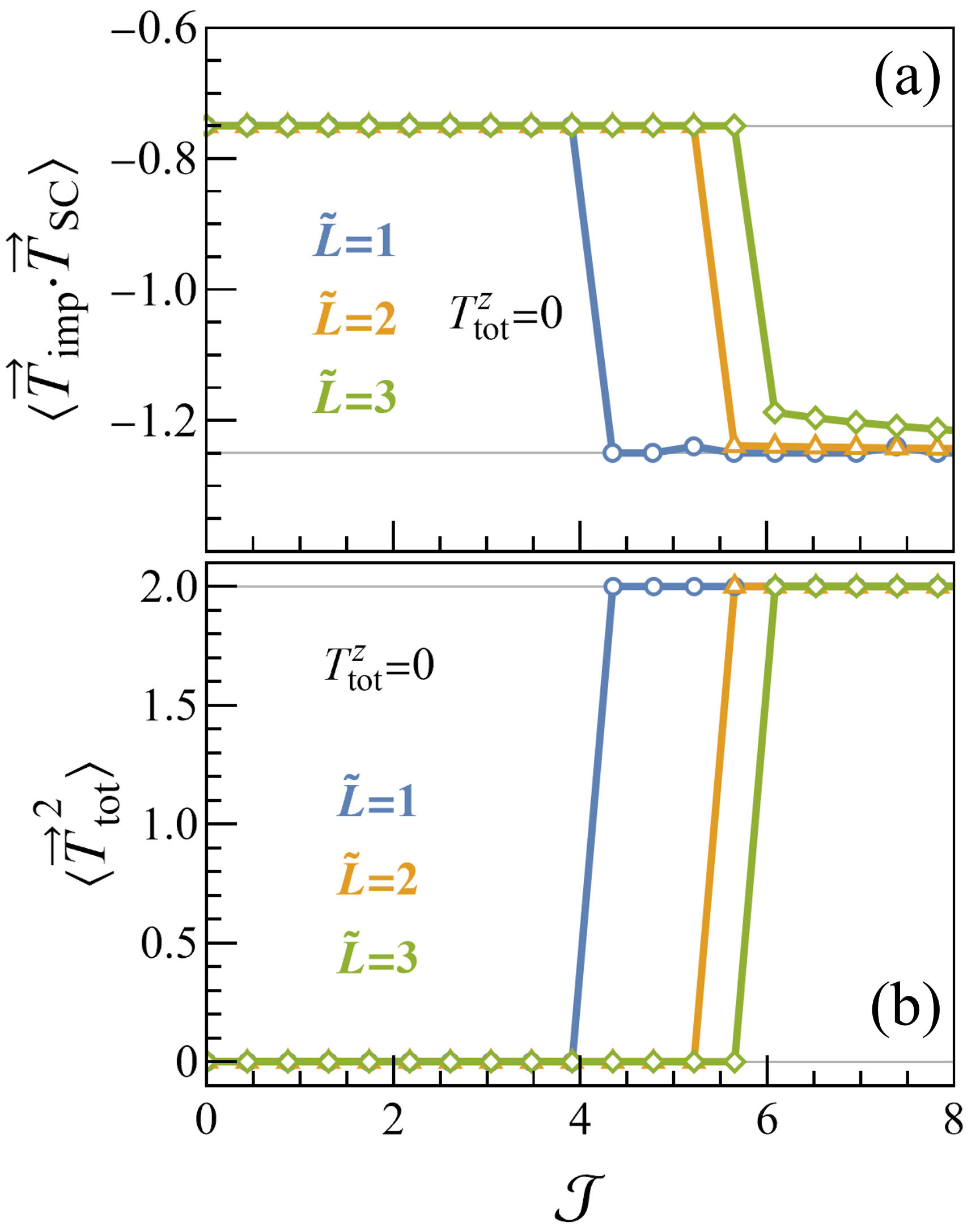}
    \caption{Color pseudo-spin properties of the lowest $T^z_{\text{tot}}=0$, odd fermion-parity states vs the dimensionless color-exchange coupling $\mathcal{J}$, for the surrogate Kondo models $\tilde{L}=1,2,3$. (a) Impurity-bulk color pseudo-spin correlation function $\langle \vec{T}_{\text{imp}}\cdot \vec{T}_{\text{SC}}\rangle$, with gridlines at $-0.75$ and $-1.25$. (c) Total color pseudo-spin squared $\langle \vec{T}_{\text{tot}}^2\rangle$, with gridlines at $0$ and $2$.}
\label{fig_4}
\end{figure}

For the more complex $\tilde{L}\geq 2$ models with multiple sets of distinct quasiparticle excitations, it becomes advantageous to describe the impurity screening in terms of the color pseudo-spin quantum numbers and associated correlation functions, see Figs.~\ref{fig_3} and~\ref{fig_4}. In Figs.~\ref{fig_3}a,b the only qualitative feature that is specific to the $\tilde{L}\geq 2$ surrogates is the gradual screening of the impurity's color with increasing coupling, as opposed to the ZBW case where the lowest lying state remains unscreened for weak to moderate couplings. Beyond this first regime, all surrogates agree on the existence of a finite region with $T_{\text{tot}}=1/2$ and  $\langle \vec{T}_{\text{imp}}\cdot \vec{T}_{\text{SC}}\rangle=-1$, corresponding to an overscreened $T_{\text{SC}}=1$ state. Here, we define
\begin{equation}
   \vec{T}_{\text{tot}}=\vec{T}_{\text{imp}}+\vec{T}_{\text{SC}}~.
\end{equation}
Finally, at strong coupling all models indicate $T_{\text{tot}}=3/2$ and  $\langle \vec{T}_{\text{imp}}\cdot \vec{T}_{\text{SC}}\rangle=-1.5$, corresponding to a $T_{\text{SC}}=2$ state.

The situation in the odd-fermion parity sector is also qualitatively similar for all surrogates, with the lowest lying state switching with increasing coupling from a color singlet $T_{\text{tot}}=0$, $T_{\text{SC}}=1/2$ to an $T_{\text{tot}}=2$, $T_{\text{SC}}=3/2$ overscreened state, as seen from Fig.~\ref{fig_4}. We note that in this sector for $T^z_{\text{tot}}=0$, the impurity's average color pseudo-spin projection also vanishes, $\langle T^z_{\text{imp}}\rangle=0$.

Overall, the screening picture given by the ZBW model is qualitatively preserved by the $\tilde{L}=2,3$ models, with the notable exception of the weak coupling regime where finite-bandwidth effects are necessary to reproduce the gradual (under)screening of the impurity's color. The $\tilde{L}=2,3$ surrogates are generally in a good semi-quantitative agreement, as seen from Figs.~\ref{fig_1},~\ref{fig_3},~\ref{fig_4}.

\section{Conclusions}
\label{sec:conclusions}

We have presented a study of the subgap states induced by a quantum impurity in high-density quark matter, where color superconductivity competes with the QCD Kondo effect. We have obtained the subgap spectrum of a two-color two-flavor QCD Kondo model with minimal numerical effort by a generalization of the few-level surrogate model solver \cite{Baran_PRB2023} to Kondo systems. The main qualitative features of the phase diagram are already present in the crudest one-level (zero-bandwidth) approximation of the superconducting bulk, and good quantitative convergence is achieved with the inclusion of one or two suitably chosen extra levels.

Concretely, we have found that with increasing color-exchange coupling the impurity experiences two phase transitions. At a moderate coupling, we switch from an unscreened/underscreened impurity to the phase where its color is screened by one of the four spin-flavor light-quark species. This gets directly superseded at a larger coupling by a phase where all four species are involved in the screening process. The alternative screened phases involving two or three quarks coupled to the impurity (previously hypothesized in Ref.~\onlinecite{Kanazawa2016Dec}) only appear as excitations in our subgap spectrum.

Due to its simplicity, our approach may prove instrumental in exploring more involved QCD scenarios, e.g. involving the full SU(3) color symmetry and/or the dynamical behavior of the coupling as a function
of density. 

\begin{acknowledgments}
This work was supported by a grant of the Romanian Ministry of Education and Research, Project No. 760122/31.07.2023 within PNRR-III-C9-2022-I9.
\end{acknowledgments}

\appendix*
\section{Surrogate Kondo models for standard BCS superconductors}

\begin{figure}[ht!]
\includegraphics[width=\columnwidth]{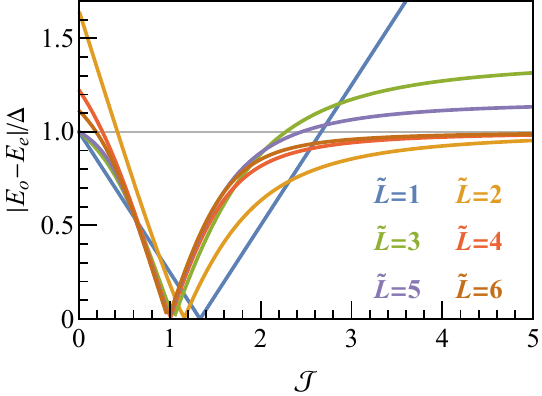}
    \caption{Energy difference $E_{o}-E_e$ (in units of $\Delta$) between the lowest even- and odd-fermion-parity states of the surrogate Kondo models of Eq.~(\ref{htilde0}) for $\tilde{L}=1,..,6$ vs the dimensionless exchange coupling $\mathcal{J}$.}
\label{fig_5}
\end{figure}

We include here further numerical results supporting the use of the prescription of Eq.~(\ref{htilde}) as a surrogate Kondo model. In particular, we consider a spin-1/2 impurity coupled to a standard BCS bulk stripped of its extra QCD-specific degrees of freedom,
\begin{equation}
\label{htilde0}
    \widetilde{\mathcal{H}}_{\text{Kondo}}=\widetilde{\mathcal{H}}_0+\widetilde{\mathcal{H}}_\text{exc}~,
\end{equation}
where
\begin{equation}
   \widetilde{\mathcal{H}}_0=\sum_{\ell=1}^{\tilde{L}}\sum_{\sigma=\uparrow\downarrow}\tilde\xi_{\ell\sigma} \tilde{c}_{\ell\sigma}^\dagger \tilde{c}_\ell-\sum_{\ell=1}^{\tilde{L}}{\Delta}(\tilde{c}_{\ell \downarrow}\tilde{c}_{\ell\uparrow} +\text{H.c.})~, 
\end{equation}
and
\begin{equation}
    \widetilde{\mathcal{H}}_\text{exc}=\mathcal{J} \,\boldsymbol{S}_{\text{imp}}\cdot\sum_{\ell,\ell'=1}^{\tilde{L}} \sqrt{\gamma_\ell \gamma_{\ell'}}\,\tilde{c}_\ell^\dagger\,\frac{\boldsymbol{\sigma}}{2}\,  \tilde{c}_{\ell'}~.
\end{equation}
This Kondo model may be obtained as a limit of the Anderson model considered in Ref.~\onlinecite{Baran_PRB2023} for a large Coulomb interaction energy $U$, which suppresses the impurity's charge fluctuations and effectively reduces it to a spin degree of freedom. In fact, the subgap spectrum shown in Fig.~\ref{fig_5} for this simple case is in excellent agreement with the full Anderson model results with $U=15\Delta$, cf. Fig. 3 of Ref.~\onlinecite{Baran_PRB2023}. In particular, the $\tilde{L}=1$ (ZBW) subgap state energy shown in Fig.~\ref{fig_5} as a blue line reads $(E_o-E_e)/\Delta=1-3\mathcal{J}/4$.

\bibliography{apssamp}

\begin{thebibliography}{59}%
\makeatletter
\providecommand \@ifxundefined [1]{%
 \@ifx{#1\undefined}
}%
\providecommand \@ifnum [1]{%
 \ifnum #1\expandafter \@firstoftwo
 \else \expandafter \@secondoftwo
 \fi
}%
\providecommand \@ifx [1]{%
 \ifx #1\expandafter \@firstoftwo
 \else \expandafter \@secondoftwo
 \fi
}%
\providecommand \natexlab [1]{#1}%
\providecommand \enquote  [1]{``#1''}%
\providecommand \bibnamefont  [1]{#1}%
\providecommand \bibfnamefont [1]{#1}%
\providecommand \citenamefont [1]{#1}%
\providecommand \href@noop [0]{\@secondoftwo}%
\providecommand \href [0]{\begingroup \@sanitize@url \@href}%
\providecommand \@href[1]{\@@startlink{#1}\@@href}%
\providecommand \@@href[1]{\endgroup#1\@@endlink}%
\providecommand \@sanitize@url [0]{\catcode `\\12\catcode `\$12\catcode
  `\&12\catcode `\#12\catcode `\^12\catcode `\_12\catcode `\%12\relax}%
\providecommand \@@startlink[1]{}%
\providecommand \@@endlink[0]{}%
\providecommand \url  [0]{\begingroup\@sanitize@url \@url }%
\providecommand \@url [1]{\endgroup\@href {#1}{\urlprefix }}%
\providecommand \urlprefix  [0]{URL }%
\providecommand \Eprint [0]{\href }%
\providecommand \doibase [0]{https://doi.org/}%
\providecommand \selectlanguage [0]{\@gobble}%
\providecommand \bibinfo  [0]{\@secondoftwo}%
\providecommand \bibfield  [0]{\@secondoftwo}%
\providecommand \translation [1]{[#1]}%
\providecommand \BibitemOpen [0]{}%
\providecommand \bibitemStop [0]{}%
\providecommand \bibitemNoStop [0]{.\EOS\space}%
\providecommand \EOS [0]{\spacefactor3000\relax}%
\providecommand \BibitemShut  [1]{\csname bibitem#1\endcsname}%
\let\auto@bib@innerbib\@empty
\bibitem [{\citenamefont {Neubert}(1994)}]{Neubert1994Sep}%
  \BibitemOpen
  \bibfield  {author} {\bibinfo {author} {\bibfnamefont {M.}~\bibnamefont
  {Neubert}},\ }\bibfield  {title} {\bibinfo {title} {{Heavy-quark symmetry}},\
  }\href {https://doi.org/10.1016/0370-1573(94)90091-4} {\bibfield  {journal}
  {\bibinfo  {journal} {Phys. Rep.}\ }\textbf {\bibinfo {volume} {245}},\
  \bibinfo {pages} {259} (\bibinfo {year} {1994})}\BibitemShut {NoStop}%
\bibitem [{\citenamefont {Neubert}(1996)}]{Neubert1996Oct}%
  \BibitemOpen
  \bibfield  {author} {\bibinfo {author} {\bibfnamefont {M.}~\bibnamefont
  {Neubert}},\ }\bibfield  {title} {\bibinfo {title} {{Heavy-Quark Effective
  Theory}},\ }\bibfield  {journal} {\bibinfo  {journal} {arXiv}\ }\href
  {https://doi.org/10.48550/arXiv.hep-ph/9610266}
  {10.48550/arXiv.hep-ph/9610266} (\bibinfo {year} {1996}),\ \Eprint
  {https://arxiv.org/abs/hep-ph/9610266} {hep-ph/9610266} \BibitemShut
  {NoStop}%
\bibitem [{\citenamefont {Manohar}\ and\ \citenamefont
  {Wise}(2000)}]{Manohar2000Mar}%
  \BibitemOpen
  \bibfield  {author} {\bibinfo {author} {\bibfnamefont {A.~V.}\ \bibnamefont
  {Manohar}}\ and\ \bibinfo {author} {\bibfnamefont {M.~B.}\ \bibnamefont
  {Wise}},\ }\href {https://doi.org/10.1017/CBO9780511529351} {\emph {\bibinfo
  {title} {{Heavy Quark Physics}}}}\ (\bibinfo  {publisher} {Cambridge
  University Press},\ \bibinfo {address} {Cambridge, England, UK},\ \bibinfo
  {year} {2000})\BibitemShut {NoStop}%
\bibitem [{\citenamefont {Hosaka}\ \emph {et~al.}(2017)\citenamefont {Hosaka},
  \citenamefont {Hyodo}, \citenamefont {Sudoh}, \citenamefont {Yamaguchi},\
  and\ \citenamefont {Yasui}}]{Hosaka2017Sep}%
  \BibitemOpen
  \bibfield  {author} {\bibinfo {author} {\bibfnamefont {A.}~\bibnamefont
  {Hosaka}}, \bibinfo {author} {\bibfnamefont {T.}~\bibnamefont {Hyodo}},
  \bibinfo {author} {\bibfnamefont {K.}~\bibnamefont {Sudoh}}, \bibinfo
  {author} {\bibfnamefont {Y.}~\bibnamefont {Yamaguchi}},\ and\ \bibinfo
  {author} {\bibfnamefont {S.}~\bibnamefont {Yasui}},\ }\bibfield  {title}
  {\bibinfo {title} {{Heavy hadrons in nuclear matter}},\ }\href
  {https://doi.org/10.1016/j.ppnp.2017.04.003} {\bibfield  {journal} {\bibinfo
  {journal} {Prog. Part. Nucl. Phys.}\ }\textbf {\bibinfo {volume} {96}},\
  \bibinfo {pages} {88} (\bibinfo {year} {2017})}\BibitemShut {NoStop}%
\bibitem [{\citenamefont {Yasui}\ and\ \citenamefont
  {Sudoh}(2013)}]{Yasui2013Jul}%
  \BibitemOpen
  \bibfield  {author} {\bibinfo {author} {\bibfnamefont {S.}~\bibnamefont
  {Yasui}}\ and\ \bibinfo {author} {\bibfnamefont {K.}~\bibnamefont {Sudoh}},\
  }\bibfield  {title} {\bibinfo {title} {{Heavy-quark dynamics for charm and
  bottom flavor on the Fermi surface at zero temperature}},\ }\href
  {https://doi.org/10.1103/PhysRevC.88.015201} {\bibfield  {journal} {\bibinfo
  {journal} {Phys. Rev. C}\ }\textbf {\bibinfo {volume} {88}},\ \bibinfo
  {pages} {015201} (\bibinfo {year} {2013})}\BibitemShut {NoStop}%
\bibitem [{\citenamefont {Hattori}\ \emph {et~al.}(2015)\citenamefont
  {Hattori}, \citenamefont {Itakura}, \citenamefont {Ozaki},\ and\
  \citenamefont {Yasui}}]{Hattori2015Sep}%
  \BibitemOpen
  \bibfield  {author} {\bibinfo {author} {\bibfnamefont {K.}~\bibnamefont
  {Hattori}}, \bibinfo {author} {\bibfnamefont {K.}~\bibnamefont {Itakura}},
  \bibinfo {author} {\bibfnamefont {S.}~\bibnamefont {Ozaki}},\ and\ \bibinfo
  {author} {\bibfnamefont {S.}~\bibnamefont {Yasui}},\ }\bibfield  {title}
  {\bibinfo {title} {{QCD Kondo effect: Quark matter with heavy-flavor
  impurities}},\ }\href {https://doi.org/10.1103/PhysRevD.92.065003} {\bibfield
   {journal} {\bibinfo  {journal} {Phys. Rev. D}\ }\textbf {\bibinfo {volume}
  {92}},\ \bibinfo {pages} {065003} (\bibinfo {year} {2015})}\BibitemShut
  {NoStop}%
\bibitem [{\citenamefont {Yasui}\ \emph {et~al.}(2017)\citenamefont {Yasui},
  \citenamefont {Suzuki},\ and\ \citenamefont {Itakura}}]{Yasui2017Jul}%
  \BibitemOpen
  \bibfield  {author} {\bibinfo {author} {\bibfnamefont {S.}~\bibnamefont
  {Yasui}}, \bibinfo {author} {\bibfnamefont {K.}~\bibnamefont {Suzuki}},\ and\
  \bibinfo {author} {\bibfnamefont {K.}~\bibnamefont {Itakura}},\ }\bibfield
  {title} {\bibinfo {title} {{Topology and stability of the Kondo phase in
  quark matter}},\ }\href {https://doi.org/10.1103/PhysRevD.96.014016}
  {\bibfield  {journal} {\bibinfo  {journal} {Phys. Rev. D}\ }\textbf {\bibinfo
  {volume} {96}},\ \bibinfo {pages} {014016} (\bibinfo {year}
  {2017})}\BibitemShut {NoStop}%
\bibitem [{\citenamefont {Yasui}(2017)}]{Yasui2017Oct}%
  \BibitemOpen
  \bibfield  {author} {\bibinfo {author} {\bibfnamefont {S.}~\bibnamefont
  {Yasui}},\ }\bibfield  {title} {\bibinfo {title} {{Kondo cloud of single
  heavy quark in cold and dense matter}},\ }\href
  {https://doi.org/10.1016/j.physletb.2017.08.066} {\bibfield  {journal}
  {\bibinfo  {journal} {Phys. Lett. B}\ }\textbf {\bibinfo {volume} {773}},\
  \bibinfo {pages} {428} (\bibinfo {year} {2017})}\BibitemShut {NoStop}%
\bibitem [{\citenamefont {Yasui}\ and\ \citenamefont
  {Ozaki}(2017)}]{Yasui2017Dec}%
  \BibitemOpen
  \bibfield  {author} {\bibinfo {author} {\bibfnamefont {S.}~\bibnamefont
  {Yasui}}\ and\ \bibinfo {author} {\bibfnamefont {S.}~\bibnamefont {Ozaki}},\
  }\bibfield  {title} {\bibinfo {title} {{Transport coefficients from the QCD
  Kondo effect}},\ }\href {https://doi.org/10.1103/PhysRevD.96.114027}
  {\bibfield  {journal} {\bibinfo  {journal} {Phys. Rev. D}\ }\textbf {\bibinfo
  {volume} {96}},\ \bibinfo {pages} {114027} (\bibinfo {year}
  {2017})}\BibitemShut {NoStop}%
\bibitem [{\citenamefont {Kimura}\ and\ \citenamefont
  {Ozaki}(2019)}]{Kimura2019Jan}%
  \BibitemOpen
  \bibfield  {author} {\bibinfo {author} {\bibfnamefont {T.}~\bibnamefont
  {Kimura}}\ and\ \bibinfo {author} {\bibfnamefont {S.}~\bibnamefont {Ozaki}},\
  }\bibfield  {title} {\bibinfo {title} {{Conformal field theory analysis of
  the QCD Kondo effect}},\ }\href {https://doi.org/10.1103/PhysRevD.99.014040}
  {\bibfield  {journal} {\bibinfo  {journal} {Phys. Rev. D}\ }\textbf {\bibinfo
  {volume} {99}},\ \bibinfo {pages} {014040} (\bibinfo {year}
  {2019})}\BibitemShut {NoStop}%
\bibitem [{\citenamefont {Hattori}\ \emph {et~al.}(2019)\citenamefont
  {Hattori}, \citenamefont {Huang},\ and\ \citenamefont
  {Pisarski}}]{Hattori2019May}%
  \BibitemOpen
  \bibfield  {author} {\bibinfo {author} {\bibfnamefont {K.}~\bibnamefont
  {Hattori}}, \bibinfo {author} {\bibfnamefont {X.-G.}\ \bibnamefont {Huang}},\
  and\ \bibinfo {author} {\bibfnamefont {R.~D.}\ \bibnamefont {Pisarski}},\
  }\bibfield  {title} {\bibinfo {title} {{Emergent QCD Kondo effect in
  two-flavor color superconducting phase}},\ }\href
  {https://doi.org/10.1103/PhysRevD.99.094044} {\bibfield  {journal} {\bibinfo
  {journal} {Phys. Rev. D}\ }\textbf {\bibinfo {volume} {99}},\ \bibinfo
  {pages} {094044} (\bibinfo {year} {2019})}\BibitemShut {NoStop}%
\bibitem [{\citenamefont {Yasui}\ and\ \citenamefont
  {Miyamoto}(2019)}]{Yasui2019Oct}%
  \BibitemOpen
  \bibfield  {author} {\bibinfo {author} {\bibfnamefont {S.}~\bibnamefont
  {Yasui}}\ and\ \bibinfo {author} {\bibfnamefont {T.}~\bibnamefont
  {Miyamoto}},\ }\bibfield  {title} {\bibinfo {title} {{Spin-isospin Kondo
  effects for ${\mathrm{\ensuremath{\Sigma}}}_{c}$ and
  ${\mathrm{\ensuremath{\Sigma}}}_{c}^{*}$ baryons and $\overline{D}$ and
  ${\overline{D}}^{*}$ mesons}},\ }\href
  {https://doi.org/10.1103/PhysRevC.100.045201} {\bibfield  {journal} {\bibinfo
   {journal} {Phys. Rev. C}\ }\textbf {\bibinfo {volume} {100}},\ \bibinfo
  {pages} {045201} (\bibinfo {year} {2019})}\BibitemShut {NoStop}%
\bibitem [{\citenamefont {Suenaga}\ \emph {et~al.}(2021)\citenamefont
  {Suenaga}, \citenamefont {Araki}, \citenamefont {Suzuki},\ and\ \citenamefont
  {Yasui}}]{Suenaga2021Mar}%
  \BibitemOpen
  \bibfield  {author} {\bibinfo {author} {\bibfnamefont {D.}~\bibnamefont
  {Suenaga}}, \bibinfo {author} {\bibfnamefont {Y.}~\bibnamefont {Araki}},
  \bibinfo {author} {\bibfnamefont {K.}~\bibnamefont {Suzuki}},\ and\ \bibinfo
  {author} {\bibfnamefont {S.}~\bibnamefont {Yasui}},\ }\bibfield  {title}
  {\bibinfo {title} {{Chiral separation effect catalyzed by heavy
  impurities}},\ }\href {https://doi.org/10.1103/PhysRevD.103.054041}
  {\bibfield  {journal} {\bibinfo  {journal} {Phys. Rev. D}\ }\textbf {\bibinfo
  {volume} {103}},\ \bibinfo {pages} {054041} (\bibinfo {year}
  {2021})}\BibitemShut {NoStop}%
\bibitem [{\citenamefont {Ishikawa}\ \emph {et~al.}(2021)\citenamefont
  {Ishikawa}, \citenamefont {Nakayama},\ and\ \citenamefont
  {Suzuki}}]{Ishikawa2021Nov}%
  \BibitemOpen
  \bibfield  {author} {\bibinfo {author} {\bibfnamefont {T.}~\bibnamefont
  {Ishikawa}}, \bibinfo {author} {\bibfnamefont {K.}~\bibnamefont {Nakayama}},\
  and\ \bibinfo {author} {\bibfnamefont {K.}~\bibnamefont {Suzuki}},\
  }\bibfield  {title} {\bibinfo {title} {{Kondo effect with Wilson fermions}},\
  }\href {https://doi.org/10.1103/PhysRevD.104.094515} {\bibfield  {journal}
  {\bibinfo  {journal} {Phys. Rev. D}\ }\textbf {\bibinfo {volume} {104}},\
  \bibinfo {pages} {094515} (\bibinfo {year} {2021})}\BibitemShut {NoStop}%
\bibitem [{\citenamefont {Suenaga}\ \emph {et~al.}(2022)\citenamefont
  {Suenaga}, \citenamefont {Araki}, \citenamefont {Suzuki},\ and\ \citenamefont
  {Yasui}}]{Suenaga2022Apr}%
  \BibitemOpen
  \bibfield  {author} {\bibinfo {author} {\bibfnamefont {D.}~\bibnamefont
  {Suenaga}}, \bibinfo {author} {\bibfnamefont {Y.}~\bibnamefont {Araki}},
  \bibinfo {author} {\bibfnamefont {K.}~\bibnamefont {Suzuki}},\ and\ \bibinfo
  {author} {\bibfnamefont {S.}~\bibnamefont {Yasui}},\ }\bibfield  {title}
  {\bibinfo {title} {{Heavy-quark spin polarization induced by the Kondo effect
  in a magnetic field}},\ }\href {https://doi.org/10.1103/PhysRevD.105.074028}
  {\bibfield  {journal} {\bibinfo  {journal} {Phys. Rev. D}\ }\textbf {\bibinfo
  {volume} {105}},\ \bibinfo {pages} {074028} (\bibinfo {year}
  {2022})}\BibitemShut {NoStop}%
\bibitem [{\citenamefont {Yasui}\ \emph {et~al.}(2024)\citenamefont {Yasui},
  \citenamefont {Suenaga},\ and\ \citenamefont {Suzuki}}]{Yasui2024May}%
  \BibitemOpen
  \bibfield  {author} {\bibinfo {author} {\bibfnamefont {S.}~\bibnamefont
  {Yasui}}, \bibinfo {author} {\bibfnamefont {D.}~\bibnamefont {Suenaga}},\
  and\ \bibinfo {author} {\bibfnamefont {K.}~\bibnamefont {Suzuki}},\
  }\bibfield  {title} {\bibinfo {title} {{QCD Kondo effect for single heavy
  quark in chiral-symmetry broken phase}},\ }\href
  {https://doi.org/10.1103/PhysRevD.109.094031} {\bibfield  {journal} {\bibinfo
   {journal} {Phys. Rev. D}\ }\textbf {\bibinfo {volume} {109}},\ \bibinfo
  {pages} {094031} (\bibinfo {year} {2024})}\BibitemShut {NoStop}%
\bibitem [{\citenamefont {Suzuki}\ \emph {et~al.}(2017)\citenamefont {Suzuki},
  \citenamefont {Yasui},\ and\ \citenamefont {Itakura}}]{Suzuki2017Dec}%
  \BibitemOpen
  \bibfield  {author} {\bibinfo {author} {\bibfnamefont {K.}~\bibnamefont
  {Suzuki}}, \bibinfo {author} {\bibfnamefont {S.}~\bibnamefont {Yasui}},\ and\
  \bibinfo {author} {\bibfnamefont {K.}~\bibnamefont {Itakura}},\ }\bibfield
  {title} {\bibinfo {title} {{Interplay between chiral symmetry breaking and
  the QCD Kondo effect}},\ }\href {https://doi.org/10.1103/PhysRevD.96.114007}
  {\bibfield  {journal} {\bibinfo  {journal} {Phys. Rev. D}\ }\textbf {\bibinfo
  {volume} {96}},\ \bibinfo {pages} {114007} (\bibinfo {year}
  {2017})}\BibitemShut {NoStop}%
\bibitem [{\citenamefont {Kanazawa}\ and\ \citenamefont
  {Uchino}(2016)}]{Kanazawa2016Dec}%
  \BibitemOpen
  \bibfield  {author} {\bibinfo {author} {\bibfnamefont {T.}~\bibnamefont
  {Kanazawa}}\ and\ \bibinfo {author} {\bibfnamefont {S.}~\bibnamefont
  {Uchino}},\ }\bibfield  {title} {\bibinfo {title} {{Overscreened Kondo
  effect, (color) superconductivity, and Shiba states in Dirac metals and quark
  matter}},\ }\href {https://doi.org/10.1103/PhysRevD.94.114005} {\bibfield
  {journal} {\bibinfo  {journal} {Phys. Rev. D}\ }\textbf {\bibinfo {volume}
  {94}},\ \bibinfo {pages} {114005} (\bibinfo {year} {2016})}\BibitemShut
  {NoStop}%
\bibitem [{\citenamefont {Rajagopal}\ and\ \citenamefont
  {Wilczek}(2001)}]{Rajagopal2001Apr}%
  \BibitemOpen
  \bibfield  {author} {\bibinfo {author} {\bibfnamefont {K.}~\bibnamefont
  {Rajagopal}}\ and\ \bibinfo {author} {\bibfnamefont {F.}~\bibnamefont
  {Wilczek}},\ }\bibfield  {title} {\bibinfo {title} {{THE CONDENSED MATTER
  PHYSICS OF QCD}},\ }in\ \href {https://doi.org/10.1142/9789812810458_0043}
  {\emph {\bibinfo {booktitle} {{At The Frontier of Particle Physics}}}}\
  (\bibinfo  {publisher} {WORLD SCIENTIFIC},\ \bibinfo {address} {Singapore},\
  \bibinfo {year} {2001})\ pp.\ \bibinfo {pages} {2061--2151}\BibitemShut
  {NoStop}%
\bibitem [{\citenamefont {Alford}\ \emph {et~al.}(2008)\citenamefont {Alford},
  \citenamefont {Schmitt}, \citenamefont {Rajagopal},\ and\ \citenamefont
  {Sch{\ifmmode\ddot{a}\else\"{a}\fi}fer}}]{Alford2008Nov}%
  \BibitemOpen
  \bibfield  {author} {\bibinfo {author} {\bibfnamefont {M.~G.}\ \bibnamefont
  {Alford}}, \bibinfo {author} {\bibfnamefont {A.}~\bibnamefont {Schmitt}},
  \bibinfo {author} {\bibfnamefont {K.}~\bibnamefont {Rajagopal}},\ and\
  \bibinfo {author} {\bibfnamefont {T.}~\bibnamefont
  {Sch{\ifmmode\ddot{a}\else\"{a}\fi}fer}},\ }\bibfield  {title} {\bibinfo
  {title} {{Color superconductivity in dense quark matter}},\ }\href
  {https://doi.org/10.1103/RevModPhys.80.1455} {\bibfield  {journal} {\bibinfo
  {journal} {Rev. Mod. Phys.}\ }\textbf {\bibinfo {volume} {80}},\ \bibinfo
  {pages} {1455} (\bibinfo {year} {2008})}\BibitemShut {NoStop}%
\bibitem [{\citenamefont {Yu}(1965)}]{Yu1965}%
  \BibitemOpen
  \bibfield  {author} {\bibinfo {author} {\bibfnamefont {L.}~\bibnamefont
  {Yu}},\ }\bibfield  {title} {\bibinfo {title} {Bound {State} in
  {Superconductors} with {Paramagnetic} {Impurities}},\ }\href
  {https://doi.org/10.7498/aps.21.75} {\bibfield  {journal} {\bibinfo
  {journal} {Acta Phys. Sin.}\ }\textbf {\bibinfo {volume} {21}},\ \bibinfo
  {pages} {75} (\bibinfo {year} {1965})}\BibitemShut {NoStop}%
\bibitem [{\citenamefont {Shiba}(1968)}]{Shiba1968}%
  \BibitemOpen
  \bibfield  {author} {\bibinfo {author} {\bibfnamefont {H.}~\bibnamefont
  {Shiba}},\ }\bibfield  {title} {\bibinfo {title} {Classical {Spins} in
  {Superconductors}},\ }\href {https://doi.org/10.1143/PTP.40.435} {\bibfield
  {journal} {\bibinfo  {journal} {Prog. Theor. Phys.}\ }\textbf {\bibinfo
  {volume} {40}},\ \bibinfo {pages} {435} (\bibinfo {year} {1968})}\BibitemShut
  {NoStop}%
\bibitem [{\citenamefont {Rusinov}(1969)}]{Rusinov1969}%
  \BibitemOpen
  \bibfield  {author} {\bibinfo {author} {\bibfnamefont {A.~I.}\ \bibnamefont
  {Rusinov}},\ }\bibfield  {title} {\bibinfo {title} {Superconductivity near a
  paramagnetic impurity},\ }\href
  {http://adsabs.harvard.edu/abs/1969JETPL...9...85R} {\bibfield  {journal}
  {\bibinfo  {journal} {JETP Lett.}\ }\textbf {\bibinfo {volume} {9}},\
  \bibinfo {pages} {85} (\bibinfo {year} {1969})},\ \bibinfo {note} {[Zh. Eksp.
  Teor. Fiz. {\bf 9}, 146 (1968)]}\BibitemShut {NoStop}%
\bibitem [{\citenamefont {Takano}\ and\ \citenamefont
  {Matayoshi}(1969)}]{Takano1969Jan}%
  \BibitemOpen
  \bibfield  {author} {\bibinfo {author} {\bibfnamefont {F.}~\bibnamefont
  {Takano}}\ and\ \bibinfo {author} {\bibfnamefont {S.}~\bibnamefont
  {Matayoshi}},\ }\bibfield  {title} {\bibinfo {title} {{Kondo Effect in
  Superconductor}},\ }\href {https://doi.org/10.1143/PTP.41.45} {\bibfield
  {journal} {\bibinfo  {journal} {Prog. Theor. Phys.}\ }\textbf {\bibinfo
  {volume} {41}},\ \bibinfo {pages} {45} (\bibinfo {year} {1969})}\BibitemShut
  {NoStop}%
\bibitem [{\citenamefont {Sakurai}(1970)}]{Sakurai1970Dec}%
  \BibitemOpen
  \bibfield  {author} {\bibinfo {author} {\bibfnamefont {A.}~\bibnamefont
  {Sakurai}},\ }\bibfield  {title} {\bibinfo {title} {{Comments on
  Superconductors with Magnetic Impurities}},\ }\href
  {https://doi.org/10.1143/PTP.44.1472} {\bibfield  {journal} {\bibinfo
  {journal} {Prog. Theor. Phys.}\ }\textbf {\bibinfo {volume} {44}},\ \bibinfo
  {pages} {1472} (\bibinfo {year} {1970})}\BibitemShut {NoStop}%
\bibitem [{\citenamefont {M{\ifmmode\ddot{u}\else\"{u}\fi}ller-Hartmann}\ and\
  \citenamefont {Zittartz}(1971)}]{Muller-Hartmann1971Feb}%
  \BibitemOpen
  \bibfield  {author} {\bibinfo {author} {\bibfnamefont {E.}~\bibnamefont
  {M{\ifmmode\ddot{u}\else\"{u}\fi}ller-Hartmann}}\ and\ \bibinfo {author}
  {\bibfnamefont {J.}~\bibnamefont {Zittartz}},\ }\bibfield  {title} {\bibinfo
  {title} {{Kondo Effect in Superconductors}},\ }\href
  {https://doi.org/10.1103/PhysRevLett.26.428} {\bibfield  {journal} {\bibinfo
  {journal} {Phys. Rev. Lett.}\ }\textbf {\bibinfo {volume} {26}},\ \bibinfo
  {pages} {428} (\bibinfo {year} {1971})}\BibitemShut {NoStop}%
\bibitem [{\citenamefont {Matsuura}(1977)}]{Matsuura1977Jun}%
  \BibitemOpen
  \bibfield  {author} {\bibinfo {author} {\bibfnamefont {T.}~\bibnamefont
  {Matsuura}},\ }\bibfield  {title} {\bibinfo {title} {{The Effects of
  Impurities on Superconductors with Kondo Effect}},\ }\href
  {https://doi.org/10.1143/PTP.57.1823} {\bibfield  {journal} {\bibinfo
  {journal} {Prog. Theor. Phys.}\ }\textbf {\bibinfo {volume} {57}},\ \bibinfo
  {pages} {1823} (\bibinfo {year} {1977})}\BibitemShut {NoStop}%
\bibitem [{\citenamefont {Cuevas}\ \emph {et~al.}(2001)\citenamefont {Cuevas},
  \citenamefont {Levy~Yeyati},\ and\ \citenamefont
  {Mart{\ifmmode\acute{\imath}\else\'{\i}\fi}n-Rodero}}]{Cuevas2001Feb}%
  \BibitemOpen
  \bibfield  {author} {\bibinfo {author} {\bibfnamefont {J.~C.}\ \bibnamefont
  {Cuevas}}, \bibinfo {author} {\bibfnamefont {A.}~\bibnamefont
  {Levy~Yeyati}},\ and\ \bibinfo {author} {\bibfnamefont {A.}~\bibnamefont
  {Mart{\ifmmode\acute{\imath}\else\'{\i}\fi}n-Rodero}},\ }\bibfield  {title}
  {\bibinfo {title} {{Kondo effect in normal-superconductor quantum dots}},\
  }\href {https://doi.org/10.1103/PhysRevB.63.094515} {\bibfield  {journal}
  {\bibinfo  {journal} {Phys. Rev. B}\ }\textbf {\bibinfo {volume} {63}},\
  \bibinfo {pages} {094515} (\bibinfo {year} {2001})}\BibitemShut {NoStop}%
\bibitem [{\citenamefont {Franke}\ \emph {et~al.}(2011)\citenamefont {Franke},
  \citenamefont {Schulze},\ and\ \citenamefont {Pascual}}]{Franke2011May}%
  \BibitemOpen
  \bibfield  {author} {\bibinfo {author} {\bibfnamefont {K.~J.}\ \bibnamefont
  {Franke}}, \bibinfo {author} {\bibfnamefont {G.}~\bibnamefont {Schulze}},\
  and\ \bibinfo {author} {\bibfnamefont {J.~I.}\ \bibnamefont {Pascual}},\
  }\bibfield  {title} {\bibinfo {title} {{Competition of Superconducting
  Phenomena and Kondo Screening at the Nanoscale}},\ }\href
  {https://doi.org/10.1126/science.1202204} {\bibfield  {journal} {\bibinfo
  {journal} {Science}\ }\textbf {\bibinfo {volume} {332}},\ \bibinfo {pages}
  {940} (\bibinfo {year} {2011})}\BibitemShut {NoStop}%
\bibitem [{\citenamefont {Zazunov}\ \emph {et~al.}(2018)\citenamefont
  {Zazunov}, \citenamefont {Plugge},\ and\ \citenamefont
  {Egger}}]{Zazunov2018Nov}%
  \BibitemOpen
  \bibfield  {author} {\bibinfo {author} {\bibfnamefont {A.}~\bibnamefont
  {Zazunov}}, \bibinfo {author} {\bibfnamefont {S.}~\bibnamefont {Plugge}},\
  and\ \bibinfo {author} {\bibfnamefont {R.}~\bibnamefont {Egger}},\ }\bibfield
   {title} {\bibinfo {title} {{Fermi-Liquid Approach for Superconducting Kondo
  Problems}},\ }\href {https://doi.org/10.1103/PhysRevLett.121.207701}
  {\bibfield  {journal} {\bibinfo  {journal} {Phys. Rev. Lett.}\ }\textbf
  {\bibinfo {volume} {121}},\ \bibinfo {pages} {207701} (\bibinfo {year}
  {2018})}\BibitemShut {NoStop}%
\bibitem [{\citenamefont {Anderson}(1970)}]{Anderson1970Dec}%
  \BibitemOpen
  \bibfield  {author} {\bibinfo {author} {\bibfnamefont {P.~W.}\ \bibnamefont
  {Anderson}},\ }\bibfield  {title} {\bibinfo {title} {{A poor man's derivation
  of scaling laws for the Kondo problem}},\ }\href
  {https://doi.org/10.1088/0022-3719/3/12/008} {\bibfield  {journal} {\bibinfo
  {journal} {J. Phys. C: Solid State Phys.}\ }\textbf {\bibinfo {volume} {3}},\
  \bibinfo {pages} {2436} (\bibinfo {year} {1970})}\BibitemShut {NoStop}%
\bibitem [{\citenamefont {Wilson}(1975)}]{Wilson1975Oct}%
  \BibitemOpen
  \bibfield  {author} {\bibinfo {author} {\bibfnamefont {K.~G.}\ \bibnamefont
  {Wilson}},\ }\bibfield  {title} {\bibinfo {title} {{The renormalization
  group: Critical phenomena and the Kondo problem}},\ }\href
  {https://doi.org/10.1103/RevModPhys.47.773} {\bibfield  {journal} {\bibinfo
  {journal} {Rev. Mod. Phys.}\ }\textbf {\bibinfo {volume} {47}},\ \bibinfo
  {pages} {773} (\bibinfo {year} {1975})}\BibitemShut {NoStop}%
\bibitem [{\citenamefont {Andrei}(1980)}]{Andrei1980Aug}%
  \BibitemOpen
  \bibfield  {author} {\bibinfo {author} {\bibfnamefont {N.}~\bibnamefont
  {Andrei}},\ }\bibfield  {title} {\bibinfo {title} {{Diagonalization of the
  Kondo Hamiltonian}},\ }\href {https://doi.org/10.1103/PhysRevLett.45.379}
  {\bibfield  {journal} {\bibinfo  {journal} {Phys. Rev. Lett.}\ }\textbf
  {\bibinfo {volume} {45}},\ \bibinfo {pages} {379} (\bibinfo {year}
  {1980})}\BibitemShut {NoStop}%
\bibitem [{\citenamefont {Vigman}(1980)}]{Vigman1980Apr}%
  \BibitemOpen
  \bibfield  {author} {\bibinfo {author} {\bibfnamefont {P.~B.}\ \bibnamefont
  {Vigman}},\ }\bibfield  {title} {\bibinfo {title} {{Exact solution of s-d
  exchange model at T = 0}},\ }\href
  {https://ui.adsabs.harvard.edu/abs/1980JETPL..31..364V/abstract} {\bibfield
  {journal} {\bibinfo  {journal} {Soviet Journal of Experimental and
  Theoretical Physics Letters}\ }\textbf {\bibinfo {volume} {31}},\ \bibinfo
  {pages} {364} (\bibinfo {year} {1980})}\BibitemShut {NoStop}%
\bibitem [{\citenamefont {Andrei}\ \emph {et~al.}(1983)\citenamefont {Andrei},
  \citenamefont {Furuya},\ and\ \citenamefont {Lowenstein}}]{Andrei1983Apr}%
  \BibitemOpen
  \bibfield  {author} {\bibinfo {author} {\bibfnamefont {N.}~\bibnamefont
  {Andrei}}, \bibinfo {author} {\bibfnamefont {K.}~\bibnamefont {Furuya}},\
  and\ \bibinfo {author} {\bibfnamefont {J.~H.}\ \bibnamefont {Lowenstein}},\
  }\bibfield  {title} {\bibinfo {title} {{Solution of the Kondo problem}},\
  }\href {https://doi.org/10.1103/RevModPhys.55.331} {\bibfield  {journal}
  {\bibinfo  {journal} {Rev. Mod. Phys.}\ }\textbf {\bibinfo {volume} {55}},\
  \bibinfo {pages} {331} (\bibinfo {year} {1983})}\BibitemShut {NoStop}%
\bibitem [{\citenamefont {Tsvelick}\ and\ \citenamefont
  {Wiegmann}(1983)}]{Tsvelick1983Jan}%
  \BibitemOpen
  \bibfield  {author} {\bibinfo {author} {\bibfnamefont {A.~M.}\ \bibnamefont
  {Tsvelick}}\ and\ \bibinfo {author} {\bibfnamefont {P.~B.}\ \bibnamefont
  {Wiegmann}},\ }\bibfield  {title} {\bibinfo {title} {{Exact results in the
  theory of magnetic alloys}},\ }\href
  {https://www.tandfonline.com/doi/abs/10.1080/00018738300101581} {\bibfield
  {journal} {\bibinfo  {journal} {Adv. Phys.}\ } (\bibinfo {year}
  {1983})}\BibitemShut {NoStop}%
\bibitem [{\citenamefont {Read}\ and\ \citenamefont
  {Newns}(1983)}]{Read1983Jun}%
  \BibitemOpen
  \bibfield  {author} {\bibinfo {author} {\bibfnamefont {N.}~\bibnamefont
  {Read}}\ and\ \bibinfo {author} {\bibfnamefont {D.~M.}\ \bibnamefont
  {Newns}},\ }\bibfield  {title} {\bibinfo {title} {{On the solution of the
  Coqblin-Schreiffer Hamiltonian by the large-N expansion technique}},\ }\href
  {https://doi.org/10.1088/0022-3719/16/17/014} {\bibfield  {journal} {\bibinfo
   {journal} {J. Phys. C: Solid State Phys.}\ }\textbf {\bibinfo {volume}
  {16}},\ \bibinfo {pages} {3273} (\bibinfo {year} {1983})}\BibitemShut
  {NoStop}%
\bibitem [{\citenamefont {Coleman}(1984)}]{Coleman1984Mar}%
  \BibitemOpen
  \bibfield  {author} {\bibinfo {author} {\bibfnamefont {P.}~\bibnamefont
  {Coleman}},\ }\bibfield  {title} {\bibinfo {title} {{New approach to the
  mixed-valence problem}},\ }\href {https://doi.org/10.1103/PhysRevB.29.3035}
  {\bibfield  {journal} {\bibinfo  {journal} {Phys. Rev. B}\ }\textbf {\bibinfo
  {volume} {29}},\ \bibinfo {pages} {3035} (\bibinfo {year}
  {1984})}\BibitemShut {NoStop}%
\bibitem [{\citenamefont {Goldhaber-Gordon}\ \emph {et~al.}(1998)\citenamefont
  {Goldhaber-Gordon}, \citenamefont {Shtrikman}, \citenamefont {Mahalu},
  \citenamefont {Abusch-Magder}, \citenamefont {Meirav},\ and\ \citenamefont
  {Kastner}}]{Goldhaber-Gordon1998Jan}%
  \BibitemOpen
  \bibfield  {author} {\bibinfo {author} {\bibfnamefont {D.}~\bibnamefont
  {Goldhaber-Gordon}}, \bibinfo {author} {\bibfnamefont {H.}~\bibnamefont
  {Shtrikman}}, \bibinfo {author} {\bibfnamefont {D.}~\bibnamefont {Mahalu}},
  \bibinfo {author} {\bibfnamefont {D.}~\bibnamefont {Abusch-Magder}}, \bibinfo
  {author} {\bibfnamefont {U.}~\bibnamefont {Meirav}},\ and\ \bibinfo {author}
  {\bibfnamefont {M.~A.}\ \bibnamefont {Kastner}},\ }\bibfield  {title}
  {\bibinfo {title} {{Kondo effect in a single-electron transistor}},\ }\href
  {https://doi.org/10.1038/34373} {\bibfield  {journal} {\bibinfo  {journal}
  {Nature}\ }\textbf {\bibinfo {volume} {391}},\ \bibinfo {pages} {156}
  (\bibinfo {year} {1998})}\BibitemShut {NoStop}%
\bibitem [{\citenamefont {Cronenwett}\ \emph {et~al.}(1998)\citenamefont
  {Cronenwett}, \citenamefont {Oosterkamp},\ and\ \citenamefont
  {Kouwenhoven}}]{Cronenwett1998Jul}%
  \BibitemOpen
  \bibfield  {author} {\bibinfo {author} {\bibfnamefont {S.~M.}\ \bibnamefont
  {Cronenwett}}, \bibinfo {author} {\bibfnamefont {T.~H.}\ \bibnamefont
  {Oosterkamp}},\ and\ \bibinfo {author} {\bibfnamefont {L.~P.}\ \bibnamefont
  {Kouwenhoven}},\ }\bibfield  {title} {\bibinfo {title} {{A Tunable Kondo
  Effect in Quantum Dots}},\ }\href
  {https://doi.org/10.1126/science.281.5376.540} {\bibfield  {journal}
  {\bibinfo  {journal} {Science}\ }\textbf {\bibinfo {volume} {281}},\ \bibinfo
  {pages} {540} (\bibinfo {year} {1998})}\BibitemShut {NoStop}%
\bibitem [{\citenamefont {Nyg{\aa}rd}\ \emph {et~al.}(2000)\citenamefont
  {Nyg{\aa}rd}, \citenamefont {Cobden},\ and\ \citenamefont
  {Lindelof}}]{Nygard2000Nov}%
  \BibitemOpen
  \bibfield  {author} {\bibinfo {author} {\bibfnamefont {J.}~\bibnamefont
  {Nyg{\aa}rd}}, \bibinfo {author} {\bibfnamefont {D.~H.}\ \bibnamefont
  {Cobden}},\ and\ \bibinfo {author} {\bibfnamefont {P.~E.}\ \bibnamefont
  {Lindelof}},\ }\bibfield  {title} {\bibinfo {title} {{Kondo physics in carbon
  nanotubes}},\ }\href {https://doi.org/10.1038/35042545} {\bibfield  {journal}
  {\bibinfo  {journal} {Nature}\ }\textbf {\bibinfo {volume} {408}},\ \bibinfo
  {pages} {342} (\bibinfo {year} {2000})}\BibitemShut {NoStop}%
\bibitem [{\citenamefont {Quay}\ \emph {et~al.}(2007)\citenamefont {Quay},
  \citenamefont {Cumings}, \citenamefont {Gamble}, \citenamefont {Picciotto},
  \citenamefont {Kataura},\ and\ \citenamefont
  {Goldhaber-Gordon}}]{Quay2007Dec}%
  \BibitemOpen
  \bibfield  {author} {\bibinfo {author} {\bibfnamefont {C.~H.~L.}\
  \bibnamefont {Quay}}, \bibinfo {author} {\bibfnamefont {J.}~\bibnamefont
  {Cumings}}, \bibinfo {author} {\bibfnamefont {S.~J.}\ \bibnamefont {Gamble}},
  \bibinfo {author} {\bibfnamefont {R.~d.}\ \bibnamefont {Picciotto}}, \bibinfo
  {author} {\bibfnamefont {H.}~\bibnamefont {Kataura}},\ and\ \bibinfo {author}
  {\bibfnamefont {D.}~\bibnamefont {Goldhaber-Gordon}},\ }\bibfield  {title}
  {\bibinfo {title} {{Magnetic field dependence of the spin-$\frac{1}{2}$ and
  spin-1 Kondo effects in a quantum dot}},\ }\href
  {https://doi.org/10.1103/PhysRevB.76.245311} {\bibfield  {journal} {\bibinfo
  {journal} {Phys. Rev. B}\ }\textbf {\bibinfo {volume} {76}},\ \bibinfo
  {pages} {245311} (\bibinfo {year} {2007})}\BibitemShut {NoStop}%
\bibitem [{\citenamefont {Jespersen}\ \emph {et~al.}(2006)\citenamefont
  {Jespersen}, \citenamefont {Aagesen}, \citenamefont {S{\o}rensen},
  \citenamefont {Lindelof},\ and\ \citenamefont
  {Nyg{\aa}rd}}]{Jespersen2006Dec}%
  \BibitemOpen
  \bibfield  {author} {\bibinfo {author} {\bibfnamefont {T.~S.}\ \bibnamefont
  {Jespersen}}, \bibinfo {author} {\bibfnamefont {M.}~\bibnamefont {Aagesen}},
  \bibinfo {author} {\bibfnamefont {C.}~\bibnamefont {S{\o}rensen}}, \bibinfo
  {author} {\bibfnamefont {P.~E.}\ \bibnamefont {Lindelof}},\ and\ \bibinfo
  {author} {\bibfnamefont {J.}~\bibnamefont {Nyg{\aa}rd}},\ }\bibfield  {title}
  {\bibinfo {title} {{Kondo physics in tunable semiconductor nanowire quantum
  dots}},\ }\href {https://doi.org/10.1103/PhysRevB.74.233304} {\bibfield
  {journal} {\bibinfo  {journal} {Phys. Rev. B}\ }\textbf {\bibinfo {volume}
  {74}},\ \bibinfo {pages} {233304} (\bibinfo {year} {2006})}\BibitemShut
  {NoStop}%
\bibitem [{\citenamefont {Csonka}\ \emph {et~al.}(2008)\citenamefont {Csonka},
  \citenamefont {Hofstetter}, \citenamefont {Freitag}, \citenamefont
  {Oberholzer}, \citenamefont {Sch{\ifmmode\ddot{o}\else\"{o}\fi}nenberger},
  \citenamefont {Jespersen}, \citenamefont {Aagesen},\ and\ \citenamefont
  {Nyg{\aa}rd}}]{Csonka2008Nov}%
  \BibitemOpen
  \bibfield  {author} {\bibinfo {author} {\bibfnamefont {S.}~\bibnamefont
  {Csonka}}, \bibinfo {author} {\bibfnamefont {L.}~\bibnamefont {Hofstetter}},
  \bibinfo {author} {\bibfnamefont {F.}~\bibnamefont {Freitag}}, \bibinfo
  {author} {\bibfnamefont {S.}~\bibnamefont {Oberholzer}}, \bibinfo {author}
  {\bibfnamefont {C.}~\bibnamefont
  {Sch{\ifmmode\ddot{o}\else\"{o}\fi}nenberger}}, \bibinfo {author}
  {\bibfnamefont {T.~S.}\ \bibnamefont {Jespersen}}, \bibinfo {author}
  {\bibfnamefont {M.}~\bibnamefont {Aagesen}},\ and\ \bibinfo {author}
  {\bibfnamefont {J.}~\bibnamefont {Nyg{\aa}rd}},\ }\bibfield  {title}
  {\bibinfo {title} {{Giant Fluctuations and Gate Control of the g-Factor in
  InAs Nanowire Quantum Dots}},\ }\href {https://doi.org/10.1021/nl802418w}
  {\bibfield  {journal} {\bibinfo  {journal} {Nano Lett.}\ }\textbf {\bibinfo
  {volume} {8}},\ \bibinfo {pages} {3932} (\bibinfo {year} {2008})}\BibitemShut
  {NoStop}%
\bibitem [{\citenamefont {Bauer}\ \emph {et~al.}(2007)\citenamefont {Bauer},
  \citenamefont {Oguri},\ and\ \citenamefont {Hewson}}]{Bauer2007Nov}%
  \BibitemOpen
  \bibfield  {author} {\bibinfo {author} {\bibfnamefont {J.}~\bibnamefont
  {Bauer}}, \bibinfo {author} {\bibfnamefont {A.}~\bibnamefont {Oguri}},\ and\
  \bibinfo {author} {\bibfnamefont {A.~C.}\ \bibnamefont {Hewson}},\ }\bibfield
   {title} {\bibinfo {title} {{Spectral properties of locally correlated
  electrons in a Bardeen{\textendash}Cooper{\textendash}Schrieffer}},\ }\href
  {https://doi.org/10.1088/0953-8984/19/48/486211} {\bibfield  {journal}
  {\bibinfo  {journal} {J. Phys.: Condens. Matter}\ }\textbf {\bibinfo {volume}
  {19}},\ \bibinfo {pages} {486211} (\bibinfo {year} {2007})}\BibitemShut
  {NoStop}%
\bibitem [{\citenamefont {Meng}\ \emph {et~al.}(2009)\citenamefont {Meng},
  \citenamefont {Florens},\ and\ \citenamefont {Simon}}]{Meng2009Jun}%
  \BibitemOpen
  \bibfield  {author} {\bibinfo {author} {\bibfnamefont {T.}~\bibnamefont
  {Meng}}, \bibinfo {author} {\bibfnamefont {S.}~\bibnamefont {Florens}},\ and\
  \bibinfo {author} {\bibfnamefont {P.}~\bibnamefont {Simon}},\ }\bibfield
  {title} {\bibinfo {title} {{Self-consistent description of Andreev bound
  states in Josephson quantum dot devices}},\ }\href
  {https://doi.org/10.1103/PhysRevB.79.224521} {\bibfield  {journal} {\bibinfo
  {journal} {Phys. Rev. B}\ }\textbf {\bibinfo {volume} {79}},\ \bibinfo
  {pages} {224521} (\bibinfo {year} {2009})}\BibitemShut {NoStop}%
\bibitem [{\citenamefont {Oguri}\ \emph {et~al.}(2013)\citenamefont {Oguri},
  \citenamefont {Tanaka},\ and\ \citenamefont {Bauer}}]{Oguri2013}%
  \BibitemOpen
  \bibfield  {author} {\bibinfo {author} {\bibfnamefont {A.}~\bibnamefont
  {Oguri}}, \bibinfo {author} {\bibfnamefont {Y.}~\bibnamefont {Tanaka}},\ and\
  \bibinfo {author} {\bibfnamefont {J.}~\bibnamefont {Bauer}},\ }\bibfield
  {title} {\bibinfo {title} {{Interplay between Kondo and Andreev-Josephson
  effects in a quantum dot coupled to one normal and two superconducting
  leads}},\ }\href {https://doi.org/10.1103/PhysRevB.87.075432} {\bibfield
  {journal} {\bibinfo  {journal} {Phys. Rev. B}\ }\textbf {\bibinfo {volume}
  {87}},\ \bibinfo {pages} {075432} (\bibinfo {year} {2013})}\BibitemShut
  {NoStop}%
\bibitem [{\citenamefont {Kir\ifmmode~\check{s}\else \v{s}\fi{}anskas}\ \emph
  {et~al.}(2015)\citenamefont {Kir\ifmmode~\check{s}\else \v{s}\fi{}anskas},
  \citenamefont {Goldstein}, \citenamefont {Flensberg}, \citenamefont
  {Glazman},\ and\ \citenamefont {Paaske}}]{Kirsanskas2015}%
  \BibitemOpen
  \bibfield  {author} {\bibinfo {author} {\bibfnamefont {G.}~\bibnamefont
  {Kir\ifmmode~\check{s}\else \v{s}\fi{}anskas}}, \bibinfo {author}
  {\bibfnamefont {M.}~\bibnamefont {Goldstein}}, \bibinfo {author}
  {\bibfnamefont {K.}~\bibnamefont {Flensberg}}, \bibinfo {author}
  {\bibfnamefont {L.~I.}\ \bibnamefont {Glazman}},\ and\ \bibinfo {author}
  {\bibfnamefont {J.}~\bibnamefont {Paaske}},\ }\bibfield  {title} {\bibinfo
  {title} {Yu-shiba-rusinov states in phase-biased superconductor-quantum
  dot-superconductor junctions},\ }\href
  {https://doi.org/10.1103/PhysRevB.92.235422} {\bibfield  {journal} {\bibinfo
  {journal} {Phys. Rev. B}\ }\textbf {\bibinfo {volume} {92}},\ \bibinfo
  {pages} {235422} (\bibinfo {year} {2015})}\BibitemShut {NoStop}%
\bibitem [{\citenamefont {Baran}\ \emph {et~al.}(2023)\citenamefont {Baran},
  \citenamefont {Frost},\ and\ \citenamefont {Paaske}}]{Baran_PRB2023}%
  \BibitemOpen
  \bibfield  {author} {\bibinfo {author} {\bibfnamefont {V.~V.}\ \bibnamefont
  {Baran}}, \bibinfo {author} {\bibfnamefont {E.~J.~P.}\ \bibnamefont
  {Frost}},\ and\ \bibinfo {author} {\bibfnamefont {J.}~\bibnamefont
  {Paaske}},\ }\bibfield  {title} {\bibinfo {title} {Surrogate model solver for
  impurity-induced superconducting subgap states},\ }\href
  {https://doi.org/10.1103/PhysRevB.108.L220506} {\bibfield  {journal}
  {\bibinfo  {journal} {Phys. Rev. B}\ }\textbf {\bibinfo {volume} {108}},\
  \bibinfo {pages} {L220506} (\bibinfo {year} {2023})}\BibitemShut {NoStop}%
\bibitem [{\citenamefont {Baran}\ and\ \citenamefont
  {Paaske}(2024{\natexlab{a}})}]{Baran2024Jun}%
  \BibitemOpen
  \bibfield  {author} {\bibinfo {author} {\bibfnamefont {V.~V.}\ \bibnamefont
  {Baran}}\ and\ \bibinfo {author} {\bibfnamefont {J.}~\bibnamefont {Paaske}},\
  }\bibfield  {title} {\bibinfo {title} {{BCS surrogate models for floating
  superconductor-semiconductor hybrids}},\ }\href
  {https://doi.org/10.1103/PhysRevB.109.224501} {\bibfield  {journal} {\bibinfo
   {journal} {Phys. Rev. B}\ }\textbf {\bibinfo {volume} {109}},\ \bibinfo
  {pages} {224501} (\bibinfo {year} {2024}{\natexlab{a}})}\BibitemShut
  {NoStop}%
\bibitem [{\citenamefont {Baran}\ and\ \citenamefont
  {Paaske}(2024{\natexlab{b}})}]{Baran2024Aug}%
  \BibitemOpen
  \bibfield  {author} {\bibinfo {author} {\bibfnamefont {V.~V.}\ \bibnamefont
  {Baran}}\ and\ \bibinfo {author} {\bibfnamefont {J.}~\bibnamefont {Paaske}},\
  }\bibfield  {title} {\bibinfo {title} {{Spin-1 Haldane chains of
  superconductor-semiconductor hybrids}},\ }\href
  {https://doi.org/10.1103/PhysRevB.110.064503} {\bibfield  {journal} {\bibinfo
   {journal} {Phys. Rev. B}\ }\textbf {\bibinfo {volume} {110}},\ \bibinfo
  {pages} {064503} (\bibinfo {year} {2024}{\natexlab{b}})}\BibitemShut
  {NoStop}%
\bibitem [{\citenamefont {Kattel}\ \emph {et~al.}(2024)\citenamefont {Kattel},
  \citenamefont {Zhakenov},\ and\ \citenamefont {Andrei}}]{Kattel2024Dec}%
  \BibitemOpen
  \bibfield  {author} {\bibinfo {author} {\bibfnamefont {P.}~\bibnamefont
  {Kattel}}, \bibinfo {author} {\bibfnamefont {A.}~\bibnamefont {Zhakenov}},\
  and\ \bibinfo {author} {\bibfnamefont {N.}~\bibnamefont {Andrei}},\
  }\bibfield  {title} {\bibinfo {title} {{Overscreened spin-$\frac{1}{2}$ Kondo
  impurity and Shiba state at the edge of a one-dimensional spin-1
  superconducting wire}},\ }\bibfield  {journal} {\bibinfo  {journal} {arXiv}\
  }\href {https://doi.org/10.48550/arXiv.2412.01924}
  {10.48550/arXiv.2412.01924} (\bibinfo {year} {2024}),\ \Eprint
  {https://arxiv.org/abs/2412.01924} {2412.01924} \BibitemShut {NoStop}%
\bibitem [{\citenamefont {Alford}\ \emph {et~al.}(1998)\citenamefont {Alford},
  \citenamefont {Rajagopal},\ and\ \citenamefont {Wilczek}}]{Alford1998Mar}%
  \BibitemOpen
  \bibfield  {author} {\bibinfo {author} {\bibfnamefont {M.}~\bibnamefont
  {Alford}}, \bibinfo {author} {\bibfnamefont {K.}~\bibnamefont {Rajagopal}},\
  and\ \bibinfo {author} {\bibfnamefont {F.}~\bibnamefont {Wilczek}},\
  }\bibfield  {title} {\bibinfo {title} {{QCD at finite baryon density: nucleon
  droplets and color superconductivity}},\ }\href
  {https://doi.org/10.1016/S0370-2693(98)00051-3} {\bibfield  {journal}
  {\bibinfo  {journal} {Phys. Lett. B}\ }\textbf {\bibinfo {volume} {422}},\
  \bibinfo {pages} {247} (\bibinfo {year} {1998})}\BibitemShut {NoStop}%
\bibitem [{\citenamefont {Rapp}\ \emph {et~al.}(1998)\citenamefont {Rapp},
  \citenamefont {Sch{\ifmmode\ddot{a}\else\"{a}\fi}fer}, \citenamefont
  {Shuryak},\ and\ \citenamefont {Velkovsky}}]{Rapp1998Jul}%
  \BibitemOpen
  \bibfield  {author} {\bibinfo {author} {\bibfnamefont {R.}~\bibnamefont
  {Rapp}}, \bibinfo {author} {\bibfnamefont {T.}~\bibnamefont
  {Sch{\ifmmode\ddot{a}\else\"{a}\fi}fer}}, \bibinfo {author} {\bibfnamefont
  {E.}~\bibnamefont {Shuryak}},\ and\ \bibinfo {author} {\bibfnamefont
  {M.}~\bibnamefont {Velkovsky}},\ }\bibfield  {title} {\bibinfo {title}
  {{Diquark Bose Condensates in High Density Matter and Instantons}},\ }\href
  {https://doi.org/10.1103/PhysRevLett.81.53} {\bibfield  {journal} {\bibinfo
  {journal} {Phys. Rev. Lett.}\ }\textbf {\bibinfo {volume} {81}},\ \bibinfo
  {pages} {53} (\bibinfo {year} {1998})}\BibitemShut {NoStop}%
\bibitem [{\citenamefont {Isgur}\ and\ \citenamefont
  {Wise}(1990)}]{Isgur1990Mar}%
  \BibitemOpen
  \bibfield  {author} {\bibinfo {author} {\bibfnamefont {N.}~\bibnamefont
  {Isgur}}\ and\ \bibinfo {author} {\bibfnamefont {M.~B.}\ \bibnamefont
  {Wise}},\ }\bibfield  {title} {\bibinfo {title} {{Weak transition form
  factors between heavy mesons}},\ }\href
  {https://doi.org/10.1016/0370-2693(90)91219-2} {\bibfield  {journal}
  {\bibinfo  {journal} {Phys. Lett. B}\ }\textbf {\bibinfo {volume} {237}},\
  \bibinfo {pages} {527} (\bibinfo {year} {1990})}\BibitemShut {NoStop}%
\bibitem [{\citenamefont {White}(1992)}]{White1992Nov}%
  \BibitemOpen
  \bibfield  {author} {\bibinfo {author} {\bibfnamefont {S.~R.}\ \bibnamefont
  {White}},\ }\bibfield  {title} {\bibinfo {title} {{Density matrix formulation
  for quantum renormalization groups}},\ }\href
  {https://doi.org/10.1103/PhysRevLett.69.2863} {\bibfield  {journal} {\bibinfo
   {journal} {Phys. Rev. Lett.}\ }\textbf {\bibinfo {volume} {69}},\ \bibinfo
  {pages} {2863} (\bibinfo {year} {1992})}\BibitemShut {NoStop}%
\bibitem [{\citenamefont
  {Schollw{\ifmmode\ddot{o}\else\"{o}\fi}ck}(2011)}]{Schollwock2011Jan}%
  \BibitemOpen
  \bibfield  {author} {\bibinfo {author} {\bibfnamefont {U.}~\bibnamefont
  {Schollw{\ifmmode\ddot{o}\else\"{o}\fi}ck}},\ }\bibfield  {title} {\bibinfo
  {title} {{The density-matrix renormalization group in the age of matrix
  product states}},\ }\href {https://doi.org/10.1016/j.aop.2010.09.012}
  {\bibfield  {journal} {\bibinfo  {journal} {Ann. Phys.}\ }\textbf {\bibinfo
  {volume} {326}},\ \bibinfo {pages} {96} (\bibinfo {year} {2011})}\BibitemShut
  {NoStop}%
\bibitem [{\citenamefont {Fishman}\ \emph
  {et~al.}(2022{\natexlab{a}})\citenamefont {Fishman}, \citenamefont {White},\
  and\ \citenamefont {Stoudenmire}}]{itensor}%
  \BibitemOpen
  \bibfield  {author} {\bibinfo {author} {\bibfnamefont {M.}~\bibnamefont
  {Fishman}}, \bibinfo {author} {\bibfnamefont {S.~R.}\ \bibnamefont {White}},\
  and\ \bibinfo {author} {\bibfnamefont {E.~M.}\ \bibnamefont {Stoudenmire}},\
  }\bibfield  {title} {\bibinfo {title} {{The ITensor Software Library for
  Tensor Network Calculations}},\ }\href
  {https://doi.org/10.21468/SciPostPhysCodeb.4} {\bibfield  {journal} {\bibinfo
   {journal} {SciPost Phys. Codebases}\ ,\ \bibinfo {pages} {4}} (\bibinfo
  {year} {2022}{\natexlab{a}})}\BibitemShut {NoStop}%
\bibitem [{\citenamefont {Fishman}\ \emph
  {et~al.}(2022{\natexlab{b}})\citenamefont {Fishman}, \citenamefont {White},\
  and\ \citenamefont {Stoudenmire}}]{itensor-r0.3}%
  \BibitemOpen
  \bibfield  {author} {\bibinfo {author} {\bibfnamefont {M.}~\bibnamefont
  {Fishman}}, \bibinfo {author} {\bibfnamefont {S.~R.}\ \bibnamefont {White}},\
  and\ \bibinfo {author} {\bibfnamefont {E.~M.}\ \bibnamefont {Stoudenmire}},\
  }\bibfield  {title} {\bibinfo {title} {{Codebase release 0.3 for ITensor}},\
  }\href {https://doi.org/10.21468/SciPostPhysCodeb.4-r0.3} {\bibfield
  {journal} {\bibinfo  {journal} {SciPost Phys. Codebases}\ ,\ \bibinfo {pages}
  {4}} (\bibinfo {year} {2022}{\natexlab{b}})}\BibitemShut {NoStop}%
\end{thebibliography}%

\end{document}